\documentclass[intlimits,twoside,a4paper]{article}

\usepackage{graphicx}

\usepackage[T2A]{fontenc}
\usepackage[cp1251]{inputenc}
\usepackage[eqsecnum]{cmpj3}

\issue{2022}{25}{1}{13703}
\doinumber{10.5488/CMP.25.13703}

\title[Resonance pressure of electromagnetic radiation on metal nanoparticle]%
{Resonance pressure of electromagnetic radiation on metal nanoparticle%
}
\author[N.~I. Grigorchuk]{N.~I. Grigorchuk}
\address{
 Bogolyubov Institute for
Theoretical Physics, National Academy of Sciences of Ukraine, \\
 14-b Metrologichna Street, Kyiv-143, Ukraine, 03143
}
\date{Received December 03, 2021, in final form February 18, 2022}

\begin{document}

\maketitle

\begin{abstract}
The influence of the electromagnetic irradiation pressure on a spheroidal
metallic nanoparticle at the frequencies close to the surface plasmon
vibrations has been considered. With the action of the radiation pressure,
the polarizability of metal nanoparticle becomes a tensor quantity.
The expressions for the resonance pressure components for the
cases of plane-polarized and circularly polarized light have been
derived. We have demonstrated that the resonance pressure can substantially
depend on the shape of a non-spherical nanoparticle
and its orientation with respect to the direction of light
propagation and the light polarization.

\end{abstract}

\section{Introduction}
The advent of lasers made the development of researches in the
field of microparticle trapping, confinement, and manipulation
possible~\cite{II}. The hot-electron pressure triggers the
anisotropic nanoparticle shape oscillations due to the thermal
expansion of the optically heated particles~\cite{PGM}.
The resonant radiation pressure on neutral particles was
investigated in~\cite{GSG}. Rawson and May observed
the angular stabilization of matter by radiation~\cite{RM}
which is familiar to trapping of small particles by
radiation pressure. In 1970, Ashkin~\cite{AA}
originally demonstrated, for the first time, the trapping
and the manipulation of a micron-sized dielectric spherical
particle in the field of two opposing laser beams.
Another work~\cite{AD} was devoted to the observation of
resonances in the radiation pressure on dielectric spheres.
The plasmon-resonance conditions for optical forces on small
particles was then considered in detail by Arias-Gonz{\'a}lez
and Nieto-Vesperinas~\cite{AN}. Latter on there was developed
a theory of optical tweezers~\cite{NBX}.
Much effort to develop the photonic force spectroscopy on
metallic nanoparticles was applied by Chaument and
coauthors~\cite{CRN}.

For the recent years, we have observed an intensive development of the
researches of the peculiarities inherent to the mechanisms of
light pressure action upon nanoparticles as well as the implication
of this action in the tasks of small particle manipulation.
Such applications meet a wide usage in biology, medicine,
and microelectronics. In the work~\cite{GJ} there was studied
the effect of laser induced angular momentum of a spheroidal
nanoparticle. The optical radiation force on a dielectric
sphere illuminated by a linearly polarized Airy light-sheet
was recently studied in~\cite{SLS}. A review of some relevant
problems can be found, e.g., in the work~\cite{SD}.

A theoretical study of the time-averaged force exerted upon a
spherical particle in a time-harmonic-varying electromagnetic
field was carried out in the work~\cite{AN}. The expression
obtained there for the force components depends on the gradient
of the electromagnetic wave intensity and on the
polarizability of the particle. The particle was considered spherical, so that
its polarizability was characterized by a scalar parameter.

In the present work, we consider metallic nanoparticles of the
spheroidal form. In this case, the particle polarizability
becomes a tensor and can rather strongly depend on the
form of the particle~\cite{TG}. Moreover, the high-frequency (optical)
conductivity~\cite{G1}, which is linked to the imaginary
part of the particle polarizability and defines its absorption,
also becomes a tensor. The dependence of the polarizability
of a metallic nanoparticle on its form becomes especially
appreciable in the infra-red range of frequencies. Under such
conditions, the expression for the components of the resonance
pressure on a nanoparticle with the action of the laser beam, would substantially
differ from those obtained in the spherical case.

\section{Formulation of the problem}

For particles, whose dimensions are considerably smaller than
the length of the electromagnetic wave, we apply the Rayleigh
approximation, i.e., the particle is considered as a dipole in
an uniform field. The force affecting such a particle equals
\begin{equation}
 {\bf{F}}=({\bf{d}}\cdot {\bf{\nabla}}){\,\bf{E}}^{(0)}+{\frac{1}{c}}
  \dot{\bf{d}} \times {\bf{B}}^{(0)},
   \label{eq1}
\end{equation}
where $\bf{d}$ is the dipole moment of the particle,
$\dot{\bf{d}}$ is the time derivative of $\bf{d}$,
${\bf{E}}^{(0)}$ is the electric field and ${\bf{B}}^{(0)}$ is the magnetic field,
and $c$ is the speed of light. All quantities in equation~(\ref{eq1})
are real. For our purpose, it is convenient to use
the complex quantities, using in equation~(\ref{eq1}) for an
arbitrary vector $(\bf V)$ the symbolic scheme
\begin{equation}
 {\bf V}\Rightarrow {\frac{1}{2}} ( {\bf V} + {\bf V}^{\ast} ).
  \label{eq2}
\end{equation}
We assue that the real and complex conjugate (with asterisk)
quantities are varied with time harmonically:
\begin{equation}
  {\bf V} = {\bf{V}}_0 \exp(- \ri \omega t),\qquad
    {\bf V}^{\ast} = {\bf{V}}_0^{\ast} \exp( \ri \omega t).
  \label{eq3}
\end{equation}
Here, $\omega$ is the frequency of the electromagnetic wave.
The time-dependence of the quantities in~(\ref{eq3}) should
determine the rapid oscillations of the force~(\ref{eq1}).
Therefore, the special case when this force is averaged over
the period $T$ of the wave is of our main interest here.

Let us now introduce the electromagnetic pressure force averaged
over this time period:
\begin{eqnarray}
 \overline{\bf{\cal P}}  &=& \frac{\overline{\bf F}}{S}=
  \frac{1}{4T}\int\limits_{-T/2}^{T/2} {\rd t}
   \Bigl\{[({\bf{d}} + {\bf{d}}\,^{\ast})\cdot {\bf{\nabla}}]\,
    ({\bf{E}} + {\bf{E}}^{\ast} )  \Bigr. \nonumber \\&+&\Bigl.
     \frac{1}{c}(\dot {\bf{d}} +
      \dot {\bf{d}}^{\ast})\times ({\bf{B}} + {\bf{B}}^{\ast}) \Bigr\},
       \label{eq4}
\end{eqnarray}
where the electric ${\bf E}$ and magnetic ${\bf B}$ fields can be
considered as normalized per unit surface area $S$ of the particle.
The second term in the integrand of expression~(\ref{eq4})
can be integrated by parts using the well-known Maxwell's equation
\begin{equation}
 -{\frac{1}{c}}{\frac{\rd{\bf{B}}}{\rd t}} = {\rm rot}{\,\bf{E}}.
    \label{eq5}
     \end{equation}
Then, instead of equation~(\ref{eq4}), we obtain
\begin{eqnarray}
 \overline{\bf{\cal P}} &=&
  {\frac{1}{4T}}{\int\limits_{-T/2}^{T/2} {\rd t}}\Bigl
   \{[({\bf{d}} + {\bf{d}} \,^{\ast}) \cdot {\bf{\nabla}}] ({\bf{E}}
    + {\bf{E}}^{\ast} ) \Bigr.\nonumber
     \\& + &\Bigl.
      ({\bf{d}} + {\bf{d}}\,^{\ast} )
       \times[{\bf{\nabla}} \times ({\bf{E}} + {\bf{E}}\,^{ \ast }) ]
        \Bigr\}.
         \label{eq6}
\end{eqnarray}
Now, taking advantage of the explicit dependence on time
[see equations~(\ref{eq3})], it is easy to carry out the integration
in equation~(\ref{eq6}) with resect to the time. The light pressure
averaged over the wave period that acts on a dipole with the dipole
moment ${\bf d}_0$ in the general case will have the form
\begin{eqnarray}
 \overline{\bf{\cal P}} &=& \frac{1}{4}\Bigl\{ ({\bf{d}}_0
  \cdot {\bf{\nabla}}){\bf{E}}_0^{\ast} + ({\bf{d}}_0^{\ast}
   \cdot {\bf{\nabla}}){\bf{E}}_0
    \Bigr.\nonumber\\& + &\Bigl.
     {\bf{d}}_0 \times [{\bf{\nabla}
      \times} {\bf{E}}_0^{\ast} ] + {\bf{d}}_0^{\ast }
       \times [{\bf{\nabla}} \times {\bf{E}}_0] \Bigr\}.
        \label{eq7}
\end{eqnarray}
Later, we will use formula~(\ref{eq7}) to calculate the resonant
pressure on a nanoparticle with irradiation by the laser beam.

Herein below, we consider a metallic nanoparticle with an
ellipsoid of revolution form. In the reference frame connected
to the principal axes of this ellipsoid, the dipole moment
of such a particle looks like~\cite{BH}
\begin{equation}
 d_{0j}={\frac{V}{4\piup}} { \frac{\varepsilon_{jj}-1}
 {1+L_j (\varepsilon_{jj}-1) }} E_{0j},
  \quad {j\ = \ x,\ y,\ z.}
   \label{eq8}
    \end{equation}
Here, $V$ is the volume of the particle,
$L_{j}$ are the depolarization factors,
\begin{equation}
  \varepsilon_{jj} = {\varepsilon}_{jj}^{\prime} +
   {\varepsilon}_{jj}^{\prime \prime} = {\varepsilon}^{\prime} +
    \ri{\frac{4\piup}{\omega}}\sigma_{jj},
     \label{eq9}
      \end{equation}
$\varepsilon^{\prime}$ is the real part
of the dielectric constant which has the form
\begin{equation}
 {\varepsilon}^{\prime} = 1-{\frac{\omega_{\rm pl}^{2}}{\omega^2}},
  \label{eq10}
   \end{equation}
$\omega_{\rm pl}$ is the plasma oscillation frequency, and $\sigma_{jj}$ are
the diagonal elements of the tensor of high-frequency (optical) conductivity.

We admit the characteristic dimension of the metallic particle to be smaller
than the mean free path of an electron in the direction of its scattering by
phonons. Having assumed such a dimension and the asymmetric form of the particle,
the conductivity becomes a tensor quantity as it was demonstrated in work~\cite{G2}.
In this case, the conductivity and, therefore, the dissipation are
influenced by both the electric field $E$ (electric absorption) and the
magnetic field $B$ (magnetic absorption) of the wave. For the case of the
ellipsoid of revolution, the following components of the tensor $\sigma
_{jj}$ are distinct from zero in the reference frame connected to the
principal axes of this ellipsoid:
\begin{equation}
 \sigma_{xx} = \sigma_{yy}\equiv \sigma_{\bot },\quad
  \sigma_{zz}\equiv \sigma_{\parallel},
   \label{eq11}
    \end{equation}
while the depolarization factors equal
  \begin{equation}
   L_{x}(e_p) = L_{y}(e_p) = {\frac{1}{2}}[1-L_{z}(e_p)]\equiv L_{\bot},
   \label{eq12}
    \end{equation}
    \begin{equation}
      L_z(e_p)\equiv L_{||}=\left\{\begin{array}{ll}
      \frac{1-e^2_p}{2e^3_p}\left(\ln{\frac{1+e_p}{1-e_p}-2e_p}\right), &
       R_{\bot} < R_{\Vert}, \\
        \frac{1+e^2_p}{e^3_p}(e_p-\arctan e_p), & R_{\bot} > R_{\Vert}.
         \end{array} \right.
          \label{eq13}
           \end{equation}
In expressions~(\ref{eq13}) the notation
 \begin{equation}
   e_p^2=\left\{
    \begin{array}{ll}
     1-R^2_{\bot}/R^2_{\Vert}, & R_{\bot} < R_{\Vert}, \\
      R^2_{\bot}/R^2_{\Vert}-1, & R_{\bot} > R_{\Vert},
       \end{array}
        \right.
         \label{eq14}
          \end{equation}
is introduced, where $R_{\parallel}$ and $R_{\perp}$ are the corresponding
semi-axes of the ellipsoid of revolution.

Introducing the vector components of the dipole moment in the form
 \begin{equation}
  d_{0i} = {\sum\limits_{j}{\alpha_{ij}E_{0j}}},
   \label{eq15}
    \end{equation}
equations~(\ref{eq8}) and~(\ref{eq11}) yield the following expressions for
nonzero components of the polarizability tensor $\alpha_{jj}$:
\begin{equation}
 \alpha_{xx}=\alpha_{yy}\equiv\alpha_{\bot} = {\frac{V}{4\piup}}{
  \frac{(\varepsilon_{\bot}-1)}{1+L_{\bot}(\varepsilon_{\bot}-1)}},
   \label{eq16}
    \end{equation}
  \begin{equation}
   \alpha_{zz} \equiv \alpha_{\parallel}  =
    {\frac{V}{4\piup}}{\frac{(\varepsilon_{_{\parallel}} -1)}{{1 +
     L_{\parallel} (\varepsilon_{_{\parallel}} - 1)}}},
      \label{eq17}
       \end{equation}
where
\begin{equation}
 \varepsilon_{\parallel} = {\varepsilon}^{\prime}+\ri{
  \frac{4\piup}{\omega}}\sigma_{_{\parallel}},\quad
   \varepsilon_{\bot} = {\varepsilon}^{\prime} + \ri{\frac{4\piup}{\omega} }
    \sigma_{\bot} .
     \label{eq18}
      \end{equation}
The expressions for $\sigma_{\perp}$ and $\sigma_{\parallel}$ under
various specific conditions are presented in work~\cite{TG}.
In particular, if the electric absorption dominates, simple analytical
expressions for the components $\sigma_{\perp}$ and $\sigma_{\parallel}$
can be obtained in the cases of strongly prolate ($R_{\parallel}\gg
R_{\perp}$) and strongly oblate ($R_{\parallel}\ll R_{\perp}$)
ellipsoids~\cite{TG}:
\begin{equation}
 \sigma_{\parallel}\approx {\frac{3}{2}}\sigma_{\bot}\approx{\frac{9\piup}
  {64}}{\frac{v_{\rm F}}{R_{\bot}}}{\frac{ne^2}{m\omega^{2}}},
   \quad (R_{\parallel}\gg R_{\bot}),
    \label{eq19}
     \end{equation}
\begin{equation}
 \sigma_{\parallel} \approx {\frac{1}{2}} \sigma_{\bot}
  \approx {\frac{9}{16}}{\frac{v_{\rm F}}{R_{\parallel}}}{
   \frac{ne^2}{m\omega^2}},
    \quad (R_{\parallel}\ll R_{\bot}) .
     \label{eq20}
      \end{equation}
Here, $v_{\rm F}$ is the Fermi velocity, $n$ is the
concentration of electrons, and $m$ is the electron mass.

For spherical particles ($R_{\parallel} = R_{\perp} = R$),
we obtain
\begin{equation}
 \sigma_{\parallel} = \sigma_{\bot} =
  {\frac{3}{4}} { \frac{v_{\rm F}}{R} }
   {\frac{ne^2}{m\omega^2} }.
    \label{eq21}
     \end{equation}
Formulae~(\ref{eq19}) and~(\ref{eq20}) are valid in
the case of high-frequency fields, when the frequency
of light is higher than the transit-time frequencies
($\omega > v_{\rm F}/R_{\bot}$, $v_{\rm F}/R_{\parallel}$).

Starting from formula~(\ref{eq15}) using equation~(\ref{eq11}),
the dipole moment can be written down for an arbitrary
coordinate system in the form
\begin{equation}
 {\bf{d}}_0 = \alpha_{\bot}{\bf{E}}_0 - (\alpha_{\bot}-
  \alpha_{\parallel})({\bf{n}}
   {\bf{E}}_0){\bf{n}}.
    \label{eq22}
     \end{equation}
Here, $\bf{n}$ is a unit vector directed along the axis of
revolution of the ellipsoid. Formulae~(\ref{eq22}) and~(\ref{eq7})
will serve as the basic ones for studying the light pressure
on a nanoparticle.

\section{Pressure caused by the action of an electromagnetic wave}

In order to obtain an explicit expression for the time-averaged pressure~(\ref{eq7}), it is necessary to establish the coordinate dependence of
the field ${\bf{E}}_0$. As the first example of such a dependence, we
take this field having a linear polarization along $x$-axis.
It looks as follows:
\begin{equation}
 {\bf{E}}_{0} = (E_{x},0,0),  \quad
  E_{x} = E_{0} \re^{-x^2/(2a^2)}\re^{\ri kz},
   \label{eq23}
    \end{equation}
where $a$ is the radius of the light beam.
Substituting expressions~(\ref{eq22}) and~(\ref{eq23})
into equations~(\ref{eq7}), we obtain the expressions for nonzero
components of the time-averaged particle pressure:
\begin{equation}
 \overline{\cal P}_{x} = - {\frac{x}{2a^2}} \left[ \vert E_0
  \vert^2{\,\rm Re}\,\alpha_{\bot} + \vert{\bf{E}}_0 {\bf{n}}
   \vert^2{\,\rm Re\,}(\alpha_{\parallel} - \alpha_{\bot} ) \right],
    \label{eq24}
     \end{equation}
\begin{equation}
 \overline{\cal P}_{z}  = {\frac{k}{2}}\left[\vert E_{0}\vert^2
  {\,\rm Im}\,\alpha_{\bot} + \vert{\bf{E}}_0 {\bf{n}}
   \vert^2 {\,\rm Im\,}(\alpha_{\parallel} - \alpha_{\bot} ) \right],
    \label{eq25}
     \end{equation}
where ${\bf E}^2_0$ is the energy density of electromagnetic field.
For the linear field polarization along $y$-axis, one must
change in equation~(\ref{eq24}) $x$ by $y$.

The real and imaginary parts
of the polarizability tensor can be written as~\cite{G2}
 \begin{equation}
   {\rm Re\,}\alpha_{\|\choose\bot} = \frac{V}{4\piup L_{\|\choose\bot}}
    \frac{\left[(1-\xi_m)\omega^2-\omega^2_{\|\choose\bot}\right]
     \left[\omega^2- \omega^2_{\|\choose\bot}\right]+
      \left[2\omega\gamma_{\|\choose\bot}\right]^2 }{\left[\omega^2-
       \omega^2_{\|\choose\bot}\right]^2 + \left[2\omega
        \gamma_{\|\choose\bot}\right]^2},
         \label{eq26}
          \end{equation}
and
\begin{equation}
 {\rm Im}\,\alpha_{\|\choose\bot}=\frac{V}{4\piup L_{\|\choose\bot}}
   \frac{2\omega^3\xi_m\gamma_{\|\choose\bot} }{\left[\omega^2-
    \omega^2_{\|\choose\bot}\right]^2+\left[2\omega\gamma_{\|\choose\bot}\right]^2},
     \label{eq27}
      \end{equation}
where we have introduced the notations
\begin{equation}
V=\frac{4}{3}\piup R_{\|} R^2_{\bot},
\end{equation}
\begin{equation}
  \xi_m = \frac{\epsilon_m}{\epsilon_m+L_{\|\choose
   \bot}-L_{\|\choose\bot}\epsilon_m},
    \label{eq28}
     \end{equation}
\begin{equation}
 \omega^2_{\|\choose\bot} = \frac{L_{\|\choose\bot}}{
  \epsilon_m+L_{\|\choose\bot}-L_{\|\choose\bot}\epsilon_m} \;\omega^2_{\rm pl},
   \label{eq29}
    \end{equation}
and
\begin{equation}
  \gamma_{\|\choose\bot}\equiv\gamma_{\|\choose\bot}(\omega) =
   \frac{2\piup L_{\|\choose\bot}}{\epsilon_m+L_{\|\choose\bot}-
    L_{\|\choose\bot}\epsilon_m} \sigma_{\|\choose\bot}(\omega)
     \label{eq30}
      \end{equation}
represents the half-width of the resonance curve for the light polarized
along ($\|$) or across ($\bot$) the revolution axis of the spheroid;
$\epsilon_m$ is the dielectric constant of the medium.

For the particle of a spherical form emersed in the medium with
$\epsilon_m=1$, in the field of the same wave, one gets
\begin{equation}
 \overline{\bf{\cal P}}_{z,\,{\rm sph}} = \frac{k}{2} \re^{-x^2/a^2}
  {E}^2_0\,\,\,{\rm Im}\,\alpha_{\rm sph} ,
   \label{eq31}
   \end{equation}
\begin{equation}
 \overline{\bf{\cal P}}_{i,\,{\rm sph}} = -\frac{x_i}{2a^2} \re^{-x^2/a^2}
  {E}^2_0 \,\,\,{\rm Re}\,\alpha_{\rm sph},
   \label{eq32}
    \end{equation}
where $x_i = x, y$ and
\begin{equation}
 {\rm Re}\,\alpha_{\rm sph} = R^3\frac{(\varepsilon'-1)
  (\varepsilon'+2)+(4\piup\sigma/\omega)^2}
   {(\varepsilon'+2)^2+(4\piup\sigma/\omega)^2},
    \label{eq33}
     \end{equation}
\begin{equation}
 {\rm Im}\,\alpha_{\rm sph} = R^3\frac{12\piup\sigma/\omega}
   {(\varepsilon'+2)^2+(4\piup\sigma/\omega)^2},
    \label{eq34}
     \end{equation}
\begin{equation}
 \sigma = \frac{3}{16\piup}\frac{v_{\rm F}}{R}
  \left(\frac{\omega_{\rm pl}}{\omega}\right)^2.
   \label{eq35}
    \end{equation}
 Here, $R$ is the radius of a spherical particle, $\sigma$
 is its high-frequency optical conductivity, and
 we take into account equation~(\ref{eq10}). Expressions~(\ref{eq26}) and~(\ref{eq27}), obtained for spheroidal
 particles, clearly transforms into the corresponding
 expressions~(\ref{eq33}) and~(\ref{eq34}) for
 spherical particles with the account of the equality
 $L_{\|}=L_{\bot}=1/3$. Then, the conductivity becomes
 a scalar quantity, specified in the form~(\ref{eq35}).
 At the plasma frequency, the real part of
 the permittivity tends to zero.

Consider now the elliptical polarized Gaussian beam:
\begin{equation}
 {\bf{E}}_0 = ( {\bf{b}}_1 + \ri{\bf{b}}_2 ) \re^{-(x^2+y^2)/2a^2} \re^{\ri kz},
  \label{eq36}
   \end{equation}
\begin{equation}
   {\bf{b}}_1 = (b_{1},0,0), \quad {\bf{b}}_2 = (0,b_2,0).
    \label{eq37}
\end{equation}
In this case, after substituting equations~(\ref{eq36}),~(\ref{eq37}),
and~(\ref{eq22}) into equation~(\ref{eq7}), we obtain
\begin{eqnarray}
 \overline{\cal P}_{i} &=& -{\frac{x}{2a^2}}\re^{-(x^2+y^2)/a^2}
  \left\{(b_1^2 + b_2^2){\,\rm Re\,}\alpha_{\bot}
   \right.\nonumber\\& + &\left.
    \left[({\bf{n}} \,{\bf{b}}_1)^2 + ({\bf{n}} \, {\bf{b}}_2)^2 \right]
     {\,\rm Re\,}(\alpha_{\parallel} - \alpha_{\bot} ) \right\},
      \label{eq38}
\end{eqnarray}
with $i=x,y$,
 \begin{eqnarray}
  \overline{\cal P}_{z} &=& \frac{k}{2}\re^{ - (x^2 + y^2)/a^2}
   \left\{(b_{1}^2 + b_2^2){\,\rm Im\,}\alpha_{ \bot}
    \right.\nonumber\\& + &\left.
     \left[({\bf{n}} \, {\bf{b}}_1)^2 + ({\bf{n}}\,
      {\bf{b}}_2)^2\right]{\,\rm Im\,}(\alpha_{\parallel} -
      \alpha_{ \bot}) \right\}.
       \label{eq39}
\end{eqnarray}
One should bear in mind that $\bf{n}$ is a unit vector directed along
the revolution axis of the ellipsoid. We see that in this case, the
light pressure depends on two angles --- between vectors $\bf{n}$
and ${\bf{b}}_1$ and between $\bf{n}$ and ${\bf{b}}_2$ vectors.

If the components of the unit vector, appearing in equation~(\ref{eq22}),
in a spherical coordinate system are represented as
\begin{equation}
 n_x = \sin\theta \cos\varphi,\quad n_y = \sin\theta
  \sin\varphi, \quad n_z = \cos\theta,
   \label{eq40}
    \end{equation}
then the products ${\bf nb}_1$ and ${\bf nb}_2$
in equations~(\ref{eq38}) and~(\ref{eq39}), respectively, become
\begin{equation}
 {\bf nb}_1 = {b}_1\sin\theta \cos\varphi,
  \quad {\bf nb}_2 = {b}_2\sin\theta \sin\varphi.
   \label{eq41}
    \end{equation}
In this case, the ratio of the average values of the light pressure
on a metal particle, having spheroidal and spherical forms
in the direction of incidence of the radiation can be written,
using equations~(\ref{eq39}) and~(\ref{eq31}), as follows
\begin{eqnarray}
 &&\frac{\overline{\cal P}_z}{\overline{\cal P}_z|_{\rm sph}}=
  \frac{\rm Im\,\alpha_{\bot}}{\rm Im\,\alpha_{\rm sph}}
    + \sin^2\theta\frac{({b}_1\cos\varphi)^2 +
    ({b}_2\sin\varphi)^2}{{b}_1^2+{b}_2^2}
     \frac{{\rm Im\,(\alpha_{\|}-
      \alpha_{\bot})}}{{\rm Im\,\alpha_{\rm sph}}}.
       \label{eq42}
        \end{eqnarray}
From equation~(\ref{eq42}), one can see that, contrary to the
particles having a spherical form (when $\alpha_{\perp}=\alpha_{\parallel}$),
the light pressure components for the nanoparticles with the ellipsoid of revolution
geometry acquire the dependence on the angle between the field direction
and the revolution axis of the ellipsoid. In addition, these components
depend on the particle's form itself, which is determined by the depolarization
factors~$L_j$ included in the diagonal components of the tensor $\alpha_{jj}$.
An analogous relation one can also obtain for the ratio of the conservative
pressure components $P_i$ if the imaginary parts of $\alpha$
in equation~(\ref{eq42}) are replaced by their real parts.

In the case of circular polarization,
${\bf{b}}_1 = {\bf{b}}_2 = {\bf{b}}$, so that
\begin{equation}
 ({\bf{n}}\, {\bf{b}}_1)^2+({\bf{n}}\, {\bf{b}}_2)^2 =
  (n_x^2+n_y^2)b^2 = (1-n_z^2)b^2.
   \label{eq43}
\end{equation}
That is, in this case, only the dependence on the angle between
the vector $\bf{n}$\ and the direction of the beam propagation survives.

Thus, similarly to the cases of plane-polarized and circularly
polarized light beams, the time-averaged light pressure that tests
a non-spherical metallic nanoparticle becomes angle-dependent.
In addition, this pressure depends on the particle's form
through the components $\alpha_{\bot}$ and $\alpha_{\|}$
of the polarization tensor; this dependence manifests itself
to the maximal extent in the infra-red range of the spectrum
(in the vicinity of the CO$_{2}$-laser frequency). For example,
taking $\omega_{\rm pl}\approx 8 \cdot 10^{15}$~s$^{-1} $  for
gold and $\omega = 2\cdot 10^{14}$~s$^{-1}$ for the CO$_2$-laser
frequency, we obtain ${\varepsilon}^{\prime}\approx -1600$.
Therefore, the combinations $L_{\bot,\parallel}(\varepsilon_{\bot,
\|}-1)$ that enter the denominators of formula
(\ref{eq16}) or~(\ref{eq17}) will be approximately equal
\begin{equation}
 L_{\parallel,\bot}(\varepsilon_{\parallel,\bot}-1)
  \approx -1600\,L_{\parallel,\bot}.
   \label{eq44}
    \end{equation}
Since the quantities $L_{\parallel}$ and $L_{\bot}$ may vary from 0 to 1
(provided that $2L_{\bot}+L_{\parallel}=1$), it is clear to what extent the
quantity~(\ref{eq24}) and, respectively, the quantities $\alpha_{\perp}$
and $\alpha_{\parallel}$ can be sensitive to the form of a metallic
particle within this range of frequencies.

\section{Light pressure at plasmon resonance}

From equations~(\ref{eq26}) and~(\ref{eq27}) it is easy to determine
the real and imaginary parts of $\alpha_{\parallel,\bot}$ at
resonance frequencies
\begin{equation}
 {\rm Re\,}\alpha_{\parallel, \bot} = \frac{V}{4\piup L_{\parallel,\bot}}, \quad
  {\rm Im\,}\alpha_{\parallel, \bot} = \frac{V}{(4\piup L_{\parallel,\bot})^2}
   \frac{\omega_{\parallel, \bot}}{\sigma_{\parallel, \bot}}.
    \label{eq45}
     \end{equation}
For a specified frequency, one can always choose a geometric
form of the particle such that it will experience a resonant
increase of the absorption with electromagnetic light pressure.
In particular, for particles of spheroidal shape, there exist
two forms of the spheroid which can resonantly absorb radiation.
The converse is also true: a nanoparticle of an arbitrary
geometric shape will absorb resonantly at least one
frequency. The higher is the degree of symmetry of the particle,
the smaller is the number of resonant frequencies that it can
absorb. For example, a spherical particle has one resonant
frequency, a spheroidal
particle has got two, and an ellipsoidal particle three.

As one can see from equation~(\ref{eq42}), the value of the ratio of
light pressures on a metallic particle depends both on the angle
of light incidence $\theta$ and on the light polarization angle $\varphi$.
Studies have shown that this ratio reaches a maximum value
at the angle of incidence equal to $\theta =\piup/2$.
Setting the angle in equation~(\ref{eq42}) equal to $\theta =\piup/2$,
let us investigate here how this relation changes for different
polarizations of the incident Gaussian beam with a change in
the shape of the nanoparticle. As an example, select the Cu
nanoparticle.

\begin{figure}[htb]
	\centerline{\includegraphics[width=9.6cm]{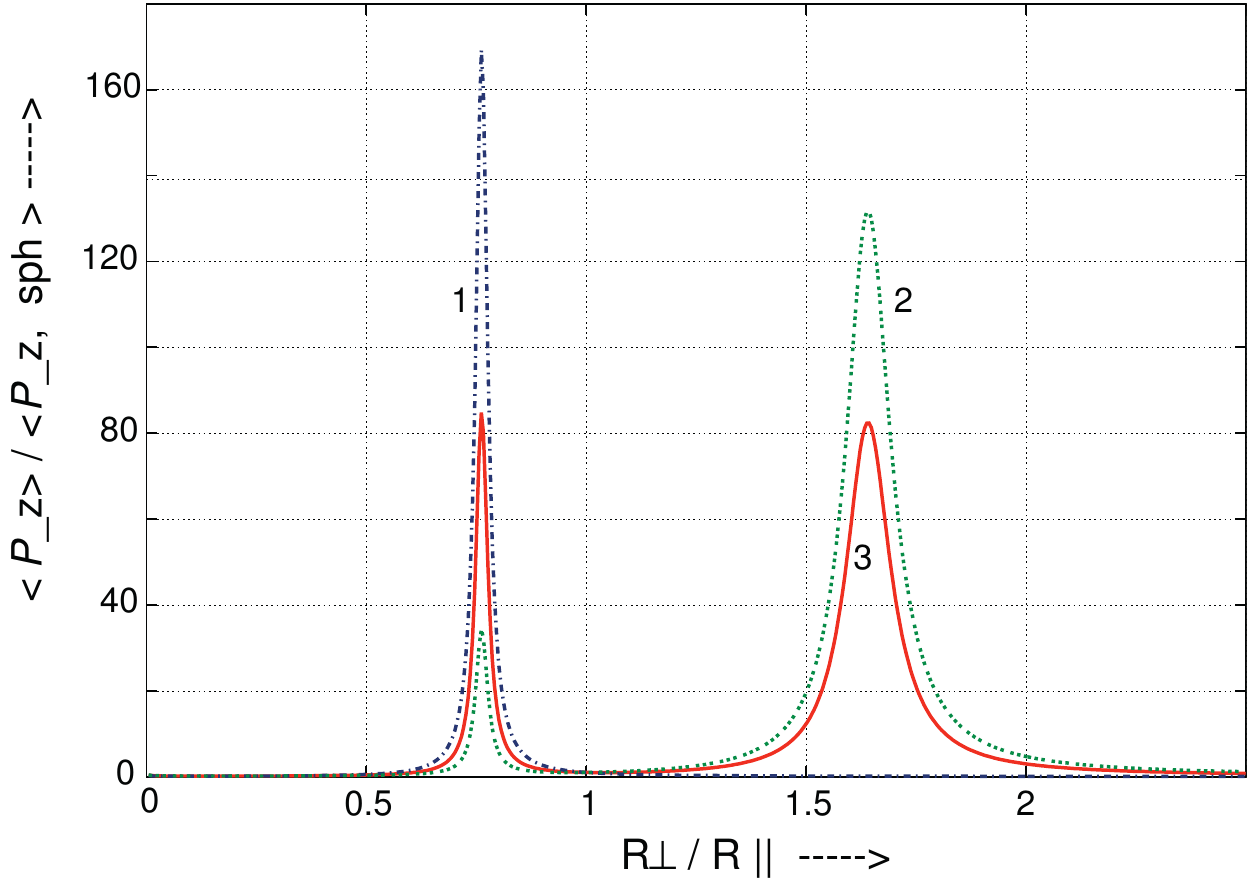}}
	\caption{(Colour online) The ratio of the light pressure
		on spheroidal Cu nanoparticle to the pressure on spherical Cu nanoparticle
		of an equal volume with the radius of $100$~\AA, as a function of
		the Cu shape, in the direction of the action of the laser beam with
		frequency $\omega\simeq 2.9\cdot 10^{15}$~s$^{-1}$ for the different
		polarization: curve~1 (short-dashed line) corresponds to the
		linear polarization; curve~2 (long-dashed line) corresponds
		to the elliptical polarization with $b_1/b_2=1/2$;
		curve~3 (solid line) corresponds to the circular polarization.} \label{figure1}
\end{figure}

\begin{figure}[htb]
	\centerline{\includegraphics[width=9.6cm]{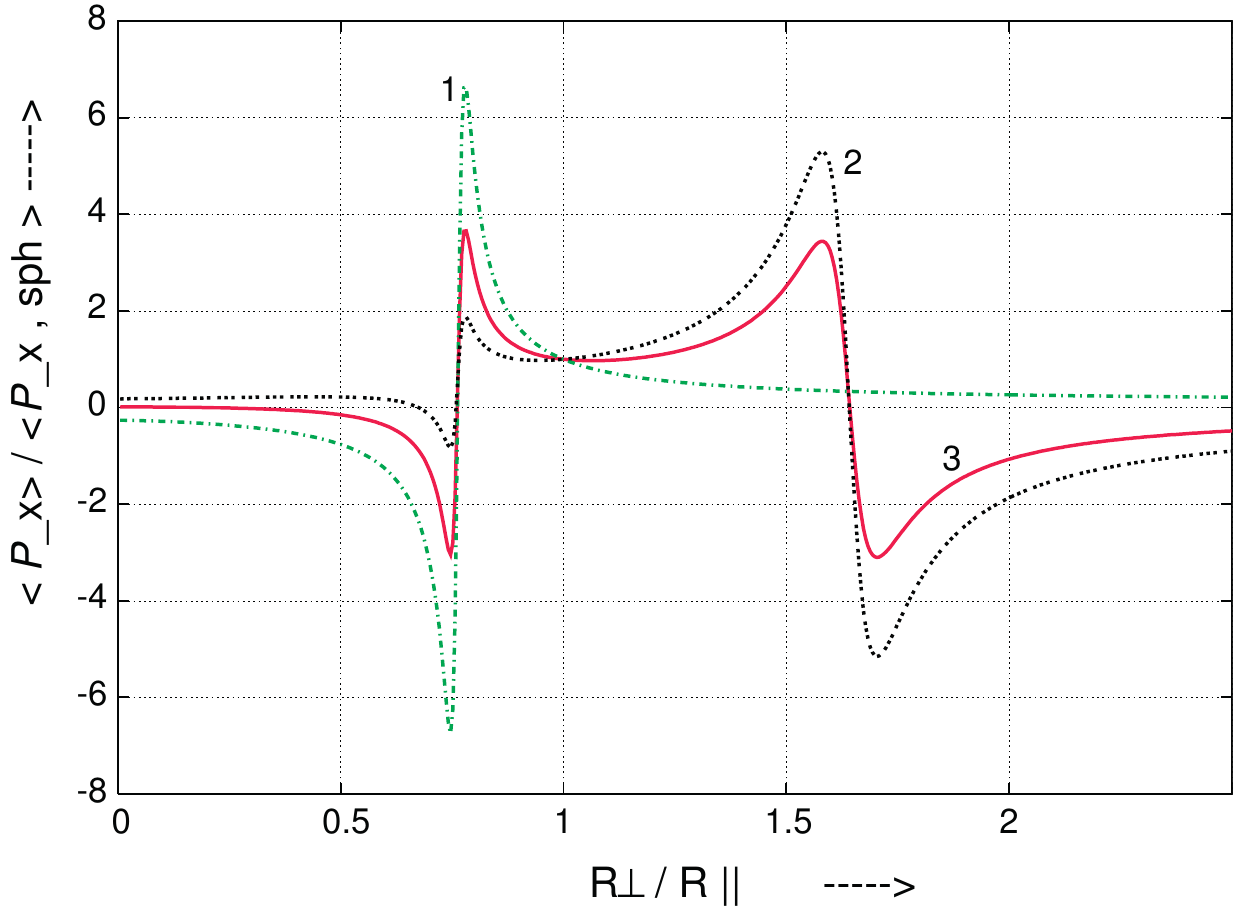}}
	\caption{(Colour online) The same as in figure~\ref{figure1}
		for the lateral $x$-direction of the action of the laser beam.} \label{figure2}
\end{figure}

Figure~\ref{figure1} illustrates the ratio of the light pressure on
spheroidal Cu nanoparticle to the light pressure on
spherical Cu nanoparticle in
the direction of incidence of the laser beam, as a
function of the deviation of the shape of the particle
from the spherical one. The frequency of laser beam was
chosen as $\omega = 2.9\cdot 10^{15}$~s$^{-1}$, which
is close to the plasmon modes in copper and $\epsilon_m=1$.
As one can see, the resonant light pressure at that frequency
is experienced by particles close to spherical shape.
Figure~\ref{figure2} shows the same dependence for the laterally directed
forces. Here and below, the force calculations are done using
equation~(\ref{eq42}) and the analogous expression obtained
with the replacement ${\rm Im}\rightarrow{\rm Re}$ for
the $i$-th pressure. As is seen in figure~\ref{figure1}, the light pressure
on the Cu particle at the plasmon resonance in the direction
of incidence of the laser beam can be hundreds of times greater
than the pressure experienced by a spherical particle of an equal volume.

In the lateral directions (figure~\ref{figure2}), this factor is considerably
smaller, not exceeding ten times. In this case, the light pressures
ratio has both positive and negative value that speaks
for the attractive or repulsive nature of the force (the
``radiation wind'') acting on a nanoparticle in this direction.
It reaches the maximum in absolute value at the angle $\varphi=\piup/2$.

With an increasing frequency of the incident radiation, the
plasmon resonance occurs in prolate nanoparticles with
a greater ratio of $R_{\bot}/R_{\|}$, while for oblate
nanoparticles it occurs with smaller values of $R_{\bot}/R_{\|}$.
In the last case, the pressure forces on the nanoparticle
fall off in absolute value.

It is also seen (figure~\ref{figure2}) that together with the resonance for
prolate particles there is also a resonance for oblate particles
in the lateral directions, and for low degrees of oblateness
this resonance is not suppressed by attenuation as it would
be the case at CO$_2$-laser frequency. In the direction of
incidence of the radiation along the $z$ axis (figure~\ref{figure1}), the
resonances appear in the form of peaks lying on each side
from the spherical shape $R_{\bot}/R_{\|}$=1.

At the pointed above plasmon frequency for Cu, the
resonance light pressure will be manifest itself [in accordance
with equations~(\ref{eq29}) and~(\ref{eq13})] for both prolate
and oblate nanoparticles with the following
values of the ratio $R_{\bot}/R_{\|}=0.763$,
and $R_{\bot}/R_{\|}=1.64$, respectively.
Consequently, the left-hand peak (figure~\ref{figure1}) and resonance
(figure~\ref{figure2}) pertain to the prolate metal particle, while the
right-hand peak and resonance belong to the oblate metal particle.

With a decrease in the angle of fall $\theta$ (with a
fixed angle $\varphi$), the peak and resonance for
the prolate~Cu nanoparticle are suppressed, whereas for
the oblate Cu nanoparticle they reach maximum values.
For a specified orientation (i.e., fixed $\theta$),
the enhancement or suppression of the peaks and
resonances in~Cu nanoparticles of different shape can
be reached by a suitable choice of polarization
of the incident radiation~\cite{KM}.

\begin{figure}[htb]
	\centerline{\includegraphics[width=9.6cm]{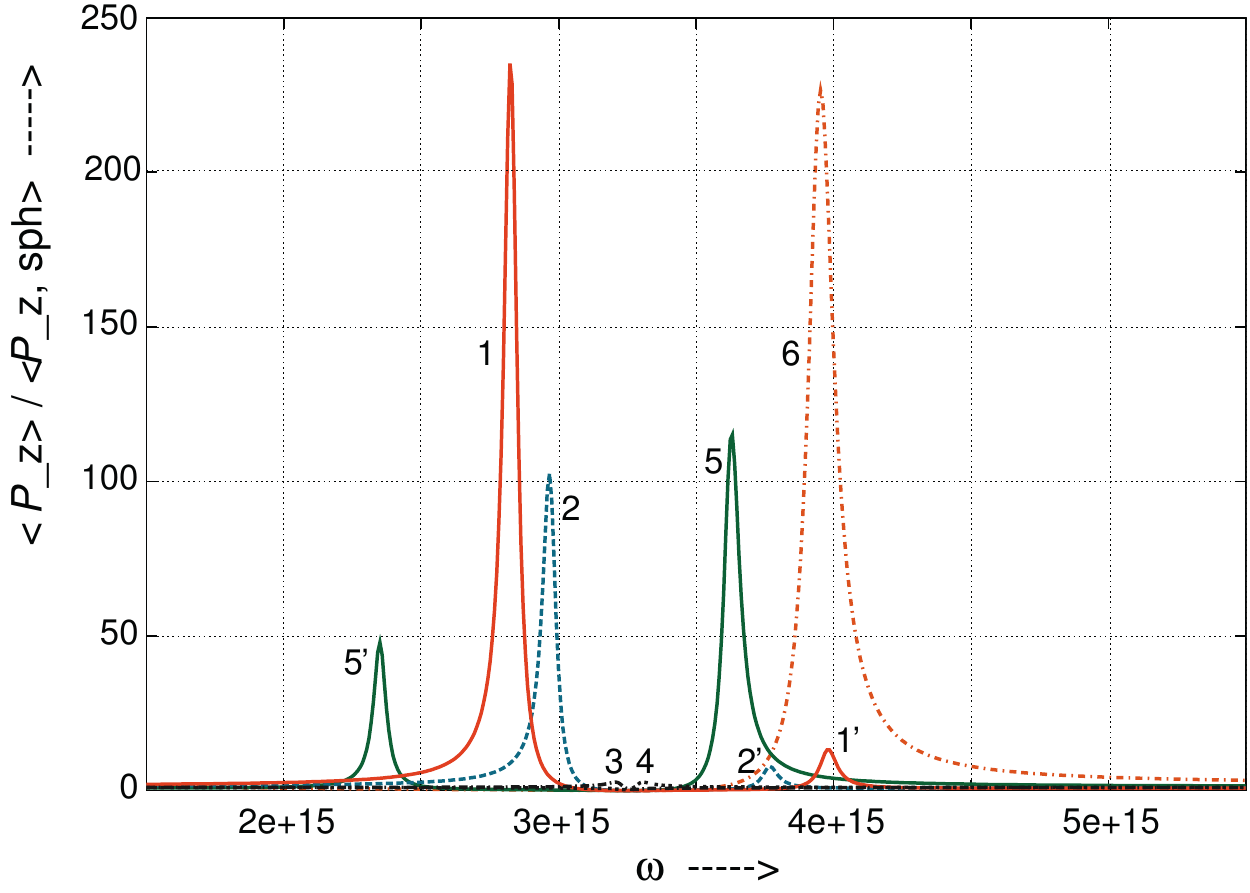}}
	\caption{(Colour online) Frequency
		dependence of the ratio of light pressure on Cu nanoparticle
		with a spheroidal shape and the light pressure on the spherical Cu
		nanoparticle of an equal volume in the direction of incident laser
		beam for nanoparticles with the radius of $100$~\AA~and different
		$R_{\bot}/R_{\|}$: 1.8~(1), 1.5~(2), 1.05~(3), 0.95~(4) 0.5~(5) 0.1~(6);
		for angles $\varphi=\piup/4$, $\theta=\piup/12$. $\epsilon_m=1.$} \label{figure3}
\end{figure}

With a change in the shape of the nanoparticle there occurs a shift
of the plasmon peaks of the light pressure. To elucidate
the nature of these shifts with the change in the degree of oblateness
or prolateness of the Cu nanoparticle, in figure~\ref{figure3} we have
plotted the frequency dependence for the ratio of the light pressure
on nanoparticles with different shapes at fixed angles.
The weak peaks 3 and 4 in this figure pertain to the Cu
nanoparticles which are close to spherical in shape.

As can be seen from figure~\ref{figure3}, the oblate nanoparticles manifest a
resonant light pressure at longer wavelengths (curves~1,~2) in
comparison with a spherical metallic nanoparticle, and
the prolate nanoparticles manifest it at shorter wavelengths (curves~5,~6).
Here, the tails of the peaks of the oblate nanoparticles extend
toward the long-wavelength side of the spectrum, while
those of the prolate nanoparticles extend toward the short-wavelength
side. As the flatness of the nanoparticle increases (curve~1),
the resonance pressure peak on a nanoparticle increases in absolute
value and shifts to the longer wavelengths, while with an increasing
elongation of the nanoparticle (curve~6) there is,
in addition to an increase of its pressure, the shift
of the peak to the shorter wavelengths of the spectrum takes place.

The peaks labeled
by numbers with a prime in figure~\ref{figure3} arise together with the
peaks labeled with the corresponding unprimed numbers; as
is seen from equation~(\ref{eq28}), this is due to the fact that they
fall into the established spectrum of $R_{\bot}/R_{\|}$ values.
Their intensity depends on two factors: the orientation of
the particle with respect to the incident radiation and/or
polarization of the radiation. In the given example, the
height of the primed peaks can be controlled by means
of the angle $\theta$; they vanish for $\theta\rightarrow 0$.

It should be noted that for prolate nanoparticles, the growth
of the intensity and the shift of the peak tend to saturation
with an increasing prolateness of the nanoparticle, after which a further
increase of the prolateness leads to a fall-off of the values
of the light pressure on a nanoparticle and the shift of the peak does
not occur. Estimates show that for $R_{\bot}/R_{\|}=1/64$,
the peak has already become rather wide,
and its shift is not very noticeable in comparison with
that for $R_{\bot}/R_{\|}=1/32$, for example.

\section{Discussions}
We have considered the radiation pressure on a very small (Rayleigh)
metal particle when the dipole approximation can be applied.
The effects of resonant radiation pressure on neutral
nanoparticles were studied previously by G{\'o}mes-Medina with
coauthors~\cite{GSG}. They showed that a small particle in a hollow
waveguide can be strongly accelerated along the guide axis while
being highly confined in a narrow zone of the cross section of the guide.
In the general case, whether a neutral nanoparticle will be attracted
to or pushed away from the high-intensity region of the laser field
and accelerated depends on the ratio of the components of the gradient
and scattering-absorbing forces~\cite{AN}.

In section 3 it was implicitly assumed that the metallic nanoparticle is placed at the center of
the beam. If one considers the deviation of metallic nanoparticle placement from the center
of the beam, then, depending on the ratio of the pressure force to
the mass of the particle, one should most likely expect one of the two scenarios:
either its revolution~(\cite{GJ}) around an axis passing through
the center of mass, or acceleration in the direction of the beam.

\section{Conclusion}

We have obtained analytical expressions for the force of resonance
pressure on a spheroidal metallic nanoparticle which is exerted by a laser
beam averaged over a period of the incident wave. It is shown that
the pressure components can substantially depend on the shape of
the particle as well as on the angles that define its orientation
both relative to the direction of the incident radiation and relative to
the polarization of the beam.

We have investigated the behavior of the electromagnetic pressure force
on a nanoparticle near plasmon resonances in spheroidal nanoparticles
in relation to the shape and orientation of the nanoparticle.
We have found the shift of the resonance peak of the pressure
towards longer wavelengths for more oblate metallic nanoparticles
and towards shorter wavelengths for more prolate ones.
We have established that the value of the light pressure with the
laser beam action on spheroidal metallic nanoparticle can differ
by orders of magnitude from the analogous light pressure acting
on a spherical metallic particle of the same volume.

\section*{Acknowledgements}
Author is grateful to the Program of the Fundamental Research of the
Department of Physics and Astronomy of the National Academy of Sciences
of Ukraine (NASU) (0121U109816) for financial support of this work.

\newpage

%

\ukrainianpart

\title{Резонансний тиск електромагнітного випромінювання на металеву частинку}
\author{М.~І. Григорчук}
\address{
	 Інститут теоретичної фізики ім.~М.~М.~Боголюбова НАН України, 
 вул.~Метрологічна, 14-б, Київ-143, Україна, 03143
}

\makeukrtitle

\begin{abstract}
\tolerance=3000%
Розглянуто вплив тиску електромагнітного випромінювання на сфероїдну металеву
частинку на частотах, близьких до частот коливань поверхневого плазмона.
Під тиском випромінювання поляризованість металевої
наночастинки стає тензорною величиною.
Одержані вирази для компонент резонансного тиску для випадків
плоско-поляризованого та циркулярно-поляризованого світла.
Показано, що резонансний тиск може істотно залежати від форми
несферичної частинки та від її орієнтації стосовно напрямку
поширення світла і його поляризації.
\keywords металеві наночастинки, плазмонний резонанс,
тиск електромагнітного випромінювання, поляризація світла
\end{abstract}


\begin{thebibliography}{20}
\bibitem{II}    Iida~T., Ishihara~H.,
                Phys. Rev. Lett., 2003, \textbf{90}, 057403,
                \doi{10.1103/PhysRevLett.90.057403}.
\bibitem{PGM}   Perner~M., Gresillon~S., M{\"a}rz~J., von~Plessen~G.,                Feldmann~J., Porstendorfer~J., Berg~K.-J., Berg~G.,
                Phys. Rev. Lett., 2000, \textbf{85}, 792--795,
                \doi{10.1103/PhysRevLett.85.792}.
\bibitem{GSG}   G{\'o}mez-Medina~R., San Jos{\'e}~P., Garc{\'i}a-Mart{\'i}n~A.,
                Lester~M., Nieto-Vesperinas~M., S{\'a}enz~J.~J.,
                Phys. Rev. Lett., 2001, \textbf{86}, 4275--4278,
               \doi{10.1103/PhysRevLett.86.4275}.
\bibitem{RM}    Rawson~E.~G., May~A.~D.,
                Appl. Phys. Lett., 1966, \textbf{8}, 93--96,
                \doi{10.1063/1.1754503}.
\bibitem{AA}    Ashkin~A., Phys. Rev. Lett., 1970, \textbf{24}, 156--159,
                \doi{10.1103/PhysRevLett.24.156}.
\bibitem{AD}    Ashkin~A., Dziedzic~J.~M.,
                Phys. Rev. Lett., 1977, \textbf{38}, 1351--1354,
                \doi{10.1103/PhysRevLett.38.1351}.
\bibitem{AN}    Arias-Gonz{\'a}lez~J.~R., Nieto-Vesperinas~M.,
                J. Opt. Soc. Am.~A, 2003, \textbf{20}, 1201--1209,\\
                \doi{10.1364/JOSAA.20.001201}.
\bibitem{NBX}   Novotny~L., Bian~R.~X., Xie~X.~S.,
                Phys. Rev. Lett., 1997, \textbf{79}, 645--648,
                \doi{10.1103/PhysRevLett.79.645}.
\bibitem{CRN}   Chaumet~P.~C., Rahmani~A., Nieto-Vesperinas~M.,
                Phys. Rev.~B., 2005, \textbf{71}, 045425,\\
                \doi{10.1103/PhysRevB.71.045425}.
\bibitem{GJ}    Grigorchuk~N.~I., 	J. Opt. Soc. Am. B, 2018, \textbf{35}, 2851--2858, \doi{10.1364/JOSAB.35.002851}.
\bibitem{SLS}   Song~N., Li~R., Sun~H., Zhang~J., Wei~B., Zhang~S., Mitri~F.~G.,
                J. Quant. Spectrosc. Radiat. Transfer, 2020, \textbf{245}, 106853,
                \doi{10.1016/j.jqsrt.2020.106853}.
\bibitem{SD}    Sukhov~S., Dogariu~A., Rep. Prog. Phys., 2017,
                \textbf{80}, 112001, \doi{10.1088/1361-6633/aa834e}.
\bibitem{TG}    Tomchuk~P.~M., Grigorchuk~N.~I.,
                Phys.~Rev.~B, 2006, \textbf{73}, 155423,
               \doi{10.1103/PhysRevB.73.155423}.
\bibitem{G1}    Grigorchuk~N.~I., 	Europhys. Lett., 2018, \textbf{121}, 67003,
                \doi{10.1209/0295-5075/121/67003}.
\bibitem{BH}    Bohren~C.~F., Huffman~D.~R.,
                Absorption and Scattering of Light by Small Particles,
                Wiley, Weinheim, 2004.
\bibitem{G2}    Grigorchuk~N.~I., Europhys. Lett., 2012, \textbf{97}, 45001,
                \doi{10.1209/0295-5075/97/45001}.
\bibitem{KM}    Kimura~H., Mann~I., J. Quant. Spectrosc. Radiat. Transfer, 1998,
                \textbf{60}, 425--438, \doi{10.1016/S0022-4073(98)00017-X}.

\end{thebibliography}
\end{document}